\documentclass[aps,prl,twocolumn,showpacs,preprintnumbers,amsmath,amssymb,floatfix]{revtex4-1}
\usepackage{amssymb}
\usepackage{graphicx}
\usepackage{dcolumn}
\usepackage{bm}
\usepackage{subfigure}
\usepackage[normalem]{ulem}
\usepackage{epstopdf}
\usepackage[colorlinks,citecolor=blue,anchorcolor=blue,linkcolor=blue,urlcolor=blue]{hyperref}
\usepackage{color}
\usepackage{multirow}
\usepackage{float}
\usepackage{titlesec}
\usepackage[utf8]{inputenc}

\begin{document}
	
\title{Anisotropic transport properties and topological Hall effect in the annealed kagome antiferromagnet FeGe}
	
\author{Jiajun Ma$^{1,*}$, Chenfei Shi$^{2,*}$, Yantao Cao$^{4,5}$, YuWei Zhang$^{1}$, Yazhou Li$^{1}$, Jiaxing Liao$^{1}$, Jialu Wang$^{1}$, Wenhe Jiao$^{3}$, Hanjie Guo$^{4}$, Chenchao Xu$^{1}$, Shixun Cao$^{2}$ Jianhui Dai$^{1}$, Jin-Ke Bao$^{1,\dag}$, Yuke Li $^{1,\ddag}$}

\affiliation{$^{1}$ School of Physics and Hangzhou Key Laboratory of Quantum Matters, Hangzhou Normal University, Hangzhou 311121, China\\
		$^{2}$ Department of Physics, Materials Genome Institute and International Center for Quantum and Molecular Structures, Shanghai University, Shanghai 200444, China\\
 $^{3}$ Key Laboratory of Quantum Precision Measurement of Zhejiang Province and Department of Applied Physics, Zhejiang University of Technology, Hangzhou 310023, China\\
 $^{4}$ Songshan Lake Materials Laboratory, Dongguan, Guangdong 523808, China \\
 $^{5}$ School of Physical Science and Technology, Lanzhou University, Lanzhou 730000, China\\
}
	
	\date{\today}

\begin{abstract}
Electron correlation often gives birth to various orders in quantum materials. Recently, a strongly correlated kagome antiferromagnet FeGe is discovered to undergo a charge density wave transition inside the A-type antiferromagnetic state, providing an opportunity to explore the interplay between charge order and magnetism. Here, we reported the observation of anisotropic resistivity and Hall effect, along with a topological Hall effect, in the annealed FeGe crystals. As the current flows along the \emph{ab}-plane, the temperature dependence of $\rho_{ab}$ exhibits a distinct resistivity loop related to a first-order transition at $T_{cdw}$. The applied magnetic fields do not alter $T_{cdw}$ but can induce a spin-flop transition at $H_{sf}$. Consequently, a field-induced large topological Hall effect is observed in the canting antiferromagnetic (CAFM) state below $T_{cant}$, which is possibly attributed to the non-trivial spin texture during the spin-flop process. Whereas, as current is parallel to \emph{c}-axis, both the field-induced transitions in $\rho_{c}$ and $\chi_{c}$ disappear. Instead, the Hall resistivity in the annealed FeGe significantly exhibits a deviation from the linear field-dependent. These findings provide valuable insight into revealing the interplay among magnetism, charge order and topology in the kagome magnets.

\textbf{Charge density wave, Hall effect, Transport properties, Kagome magnet}

\textbf{PACS number(s):}  71.45.Lr, 73.43.–f, 73.23.–b, 75.50.Ee

\end{abstract}

\maketitle

\section{I. INTRODUCTION}
	
The kagome lattice materials with corner-sharing triangular networks provide a fertile platform for the realization of correlated quantum states due to the electronic band structure of kagome lattice featuring flat band, Dirac cone and a pair of van Hove singularities \cite{PRBReview,SALi,PRBWangQh}, exemplified by magnetism \cite{FeSn,Fe3Sn2}, charge density waves (CDW) \cite{CsVSb,FeGe,PRLScVSn}, quantum spin liquid \cite{lee2007quantum}, nematicity \cite{nie2022charge} and superconductivity \cite{LaRuSi,PRBWangQh}. Recent years have witnessed the important development of kagome lattice materials, e.g., antiferromagnetic Mn$_3$Sn/Ge \cite{NMn3Sn,Mn3Ge}, magnetic Weyl-semimetal Co$_3$Sn$_2$S$_2$ \cite{MPI,wangRM,CoSnSLi}, Chern-gap magnet TbMn$_6$Sn$_6$ \cite{yinjx,JSMnSn} and CDW ordering compounds CsV$_3$Sb$_5$ and ScV$_6$Sn$_6$ \cite{CsVSb,PRLSC,PRLScVSn}, where they exhibit many interesting quantum behaviours such as anomalous/topopolocal Hall effect/Nernst effect, chiral magnetic effect and unconventional superconductivity.\par \setlength{\parskip}{0pt}
	
Recently, a strongly correlated kagome metal FeGe (B35-type) is reported to exhibit A-type collinear antiferromagnetic (AFM) order with magnetic moments along \textit{c} axis at $T_N$ = 400 K, and a short-range CDW phase at $\sim$100 K \cite{DCAFM,FeGe}. Upon further cooling to a lower temperature of 60 K, its magnetic moments form a canted double-cone AFM structure with an inter-layer turn angle for the basal-plane moment component, regardless of temperature and applied field. The cone half-angle expands as temperature decreases to $-14^\circ$ at 4.2 K\cite{DCAFM}. A key feature in FeGe is that the charge order locates deeply inside the AFM phase, leading to a significant enhancement of Fe moments \cite{DCAFM,FeGe}, which is in contrast to the CDW phase over the magnetic ordering temperature in cuprates and nickelates \cite{RevModPhys.75.1201,zhang2020intertwined}. This anomalous feature has triggered intensive investigations both experimentally and theoretically to understand the physics involving the intertwinement of AFM and CDW \cite{teng2023magnetism,miao2023signature,wu2023electron,wang2023enhanced,yin2022discovery,ma2023theory,xiangangwan,shao2023intertwining,zhang2023triple}.
	
In earlier experiments after its discovery, neutron scattering \cite{FeGe,chen2023competing}, scanning tunneling microscopy (STM) \cite{yin2022discovery,chen2023charge} and angle-resolved photoemission spectroscopy (ARPES) \cite{teng2023magnetism} supported the short-ranged 2 $\times$ 2 $\times$ 2 CDW order and a strong entanglement of charge order and magnetism in as-grown FeGe samples. Of late, long-ranged CDW order in FeGe has been achieved in the higher quality samples by post-annealing\cite{Baojinke,chen2023long,FeGe_long_aifeng}. Single crystal X-ray diffraction \cite{Baojinke,chen2023long,miao2023signature}, ARPES \cite{zhao2023photoemission} on these annealed samples indicated the dimerization of Ge1 atoms in the kagome layers along the \textit{c}-axis as the major driving force for the realization of CDW in FeGe proposed by first-principle calculations \cite{wang2023enhanced,miao2023signature}. It is also found that CDW order coexists with a pronounced edge state \cite{yin2022discovery}, pointing to a strong coupling of topology, charge order and magnetism in the kagome lattice of FeGe. However, the confliction of short-ranged and long-ranged CDW ordering implies that the as-grown FeGe samples can not really reflect the intrinsic physical properties because of the defects or other impurity phases. In addition, though the anomalous Hall effect (AHE) has been investigated in as-grown FeGe\cite{FeGe}, the origin of AHE and its anisotropy still remains unknown. Thus, the study of the intrinsic transport properties as well as the interplay between charge order and magnetism will be very urgent and necessary in the high-quality annealed FeGe crystals.\par
	
In this paper, we performed the measurements of magnetization and transport properties as current is parallel to \emph{ab}-plane and \emph{c}-direction in the annealed FeGe crystals, respectively. The anisotropic magnetization and anomalous electronic transports are observed. In-plane resistivity $\rho_{ab}$ shows a significant resistivity loop near $T_{cdw}$ associated with the first-order phase transition. In particular, $\rho_{ab}$ value reaches approximately three times larger than $\rho_c$ at room temperature, signifying an anomalous anisotropic resistivity. Applying a magnetic field on $\rho_{ab}$ and $\chi_{ab}$ can trigger a spin-flop transition at $H_{sf}$. We thus discover a field-induced large topological Hall effect in the canting AFM state below $T_{cdw}$. In addition, as \textit{H} $\parallel$ \textit{ab}-plane, \textit{I} $\parallel$ \textit{c}-axis, the Hall resistivity seriously deviates from the linear field-dependent below $T_{cdw}$. Such sharp divergency in transports can be explained by the chiral effect during the spin-flop process in the magnetic frustrated material with charge order.

\section{II. EXPERIMENTAL DETAILS }
High quality single crystals of FeGe were grown by the chemical vapor transport (CVT) method, as reported in the previous literatures \cite{FeGe,Baojinke}. The mixture of high-purity iron and germanium powders in the ratio of 1:1 with additional iodine as a transport agent was weighed and sealed in a quartz ampule under vacuum. The ampule was placed in a two-zone furnace, heated up to $600^\circ$ C with a temperature gradient of $2^\circ$ C per~$\mathrm{cm}$, and kept at this temperature profile for two weeks. The ampule was quenched in the water after the crystal growth was finished. The shiny as-grown FeGe crystals were further annealed at $370^\circ$ C under an evacuated quartz ampule for 10 days. The obtained annealed crystals of FeGe are measured and investigated in our study. The energy-dispersive X-ray (EDX) spectroscopy confirms that the actual ratio of Fe to Ge is close to 1:1 (see Figure S1 in Supplementary information (SI)).

The annealed crystal shows a hexagonal shaped with a size of $2\times1\times 1$~$\mathrm{mm}^3$. The crystal X-ray diffraction patterns help to determine the crystal orientations through a D/MaxrA diffractometer with CuK$_{\alpha}$ radiation at 300 K. The energy dispersive X-ray (EDX) spectroscopy confirms the real stoichiometry of FeGe. The annealed crystal samples are cut into two rectangular pieces with the size of $\sim 0.5 \times 1$~$\mathrm{mm}^2$ and 0.6 $\times 1.8$~$\mathrm{mm}^2$ and then polished (See the inset of Fig.S1 in SI). The (magneto) resistivity and Hall coefficient measurements were performed in the two pieces using the standard four-terminal method in a Cryogenic-14 T magnet system. Both in-plane resistivity $\rho_{ab}$ and out of plane resistivity $\rho_c$ are measured when current is parallel to \emph{ab}-plane or \emph{c}-direction. The configuration of (Hall) resistivity measurements is displayed in the inset of Figure S2 (See detailed information in SI). The positive and negative magnetic fields perpendicular to current are applied to eliminate the contribution of extrinsic resistivity due to a slight misalignment of the voltage during measurements. The magnetization measurements were performed using a commercial SQUID magnetometer.
	
\section{III. RESULTS AND DISCUSSION}
		
\begin{figure}
\includegraphics[angle=0,width=8cm,clip]{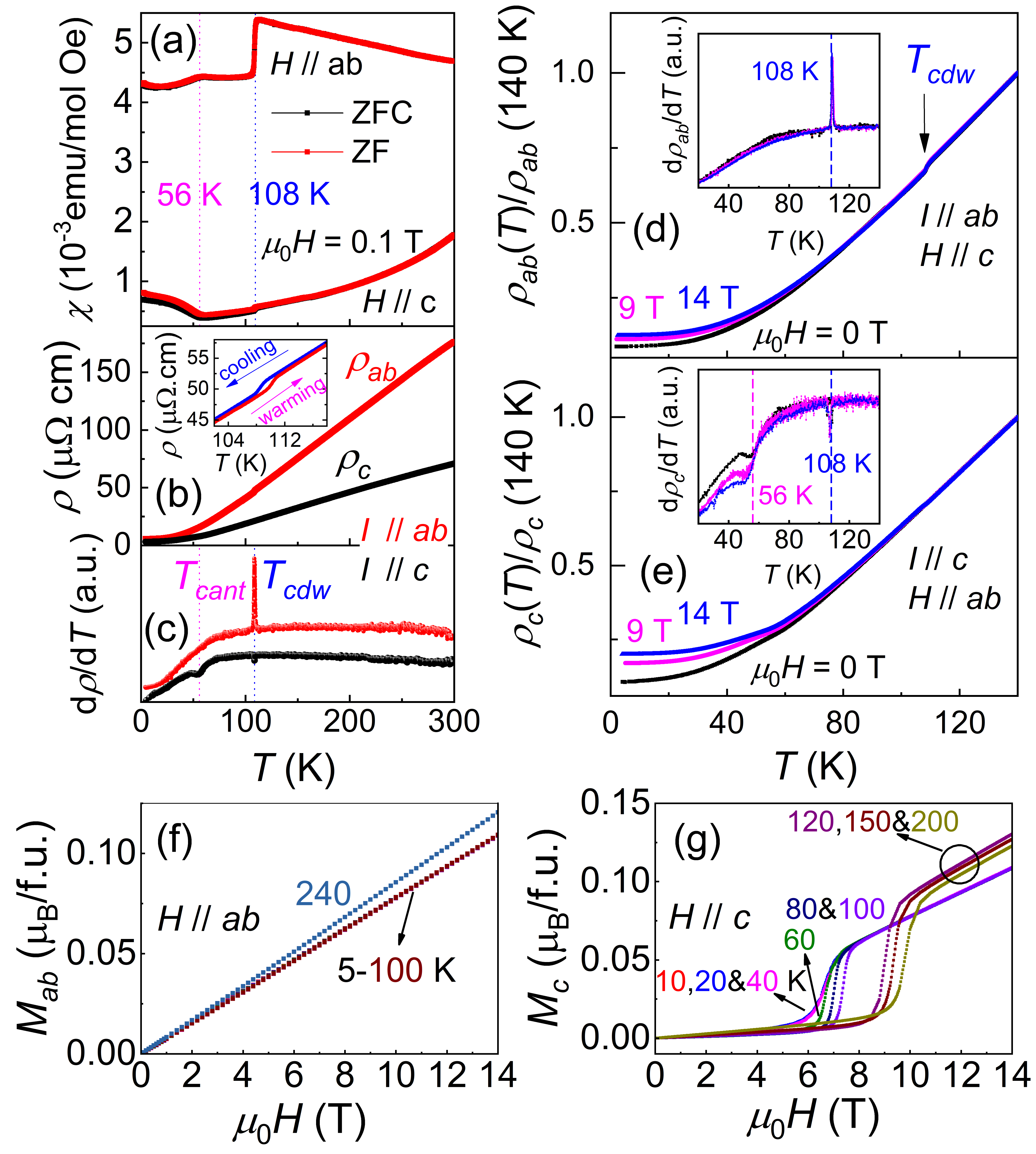}
\caption{(a) In-plane ($\chi_{ab}$) and out-of-plane ($\chi_{c}$) magnetic susceptibility as a function of temperature under 0.1 T. (b) Temperature dependence of the in-plane resistivity $\rho_{ab}$ and out-of-plane $\rho_{c}$ at zero field. The inset shows the enlarge plot around $T_{cdw}$ with cooling and warming modes. (c) The differentiation of resistivity d$\rho/$d$T$. (d) $\rho_{ab}$ and $\rho_{c}$ (e) as functions of temperature below 140 K under 9 T and 14 T. The insets show the differentiation of resistivity d$\rho/$d$T$ under zero field, 9 T and 14 T. The in-plane magnetization $M_{ab}$ ($H \parallel$ \emph{ab}-plane) (f)  and the out-plane magnetization $M_{c}$ ($H \parallel$ \emph{c}-direction)(g) as functions of magnetic fields for several representative temperatures.}
\label{1}
\end{figure}
	
FeGe crystallizes in a hexagonal structure with space group $P6/$mmm, which consists of stacks of Fe kagome planes with both in-plane and inter-plane Ge1 and Ge2 atoms along \emph{c}-axis. At $T_N =$ 410 K, it shows A-type AFM order, and its moments order ferromagnetically along the \emph{c}-axis within each plane and anti-aligned between layers, as shown in Fig. \ref{1}(a). The shining FeGe crystal with a millimeter size forms a hexagonal prism along the \emph{c}-axis. Its single crystal diffraction patterns clearly show a set of (00l) peaks without any extra impurity phase (see Figure S1(c) in SI), implying the high-quality single crystals in our measurements. The measured magnetic susceptibility $\chi_{ab}(T)$ in the annealed FeGe samples (FG-370) shows a sharp drop at 108 K related to the CDW transition in Fig. \ref{1}(a), followed by a small kink at 56 K associated with the canted AFM order with decreasing temperature as \emph{H} \textit{$\parallel$} \emph{ab}-plane. This CDW transition is more prominent compared with as-grown FeGe sample \cite{FeGe}. The two transitions are also observed in $\chi_c$, although its value is about four times smaller than $\chi_{ab}$, indicating a large anisotropic behavior in the annealed FG-370.

Correspondingly, as the current \textit{I} \textit{$\parallel$} \textit{ab}-plane, the $\rho_{ab}$ at room temperature reaches 176 $\mu \Omega$~$\mathrm{cm}$, which is about three times larger than the value of 60 $\mu \Omega$~$\mathrm{cm}$ in $\rho_c$, as shown in Fig. \ref{1}(b). The unexpectedly larger in-plane resistivity implies abnormal anisotropic resistivity behaviour, since the interlayer coupling is usually weaker than the intralayer coupling in these layered kagome materials. The unusual anisotropic resistivity behaviors can be ascribed to the lattice-derived band structures, similar to the findings in the longitudinal thermoelectric effect(not presented here). The flat band in the $k_x-k_y$ plane can substantially suppress the movement of electron due to the infinite large effective mass. In contrast, the sharp band dispersion in the $k_z$ plane facilitates the movement of electrons, leading to a smaller out-of-plane resistivity, similar to the case in the kagome magnet CoSn\cite{CoSn}. Cooling down the sample to lower temperature, $\rho_{ab}$ shows a linear decrease but suddenly drops at the CDW transition temperature of 108 K (see Fig. \ref{1}(d)), and its differentiation d$\rho_{ab}/$d$T$ also shows a sharp peak at $T_{cdw}$ (see Fig. \ref{1}(c)). More interestingly, a clear heat hysteresis in $\rho_{ab}$ near $T_{cdw}$ is at first observed in the FeGe samples, as shown in the inset of Fig. \ref{1}(b), suggesting a first-order phase transition. $\rho_c$ monotonically decreases with decreasing temperature. No anomaly at $T_{cdw}$ is observed but a small hump around $T_{cant} =$ 56 K related to canting AFM transition can be distinguished (see Fig. \ref{1}(e)). Through performing the differentiation of $\rho_c$ in Fig. \ref{1}(c), two dips at $T_{cdw}$ and $T_{cant}$ can be clearly detected respectively.

The typical values of $\rho_{ab}$ ($\rho_c$) at 2 K are found to be 6.0 (3.2) $\mu \Omega$~cm, yielding the RRR ($= \rho_{300\mathrm{K}}/\rho_{2\mathrm{K}}$) is 29.3 (21.8). These values are comparable to the topological magnetic compounds Co$_3$Sn$_2$S$_2$ \cite{CoSnSLi}, Co$_2$MnGa \cite{CoMnGa} and Fe$_3$GeTe$_2$ \cite{Lisign}, while the RRR is much lower than typical semimetals Bi \cite{PRBi} and TaSb$_2$  \cite{PRBLi}, et.al.. Applying magnetic fields of 9 T and 14 T into both \textit{ab}-plane and \textit{c}-direction in Fig. \ref{1}(d)-\ref{1}(e), we found that the increase of $\rho_{ab}$ is much smaller than that of $\rho_c$ at low temperatures. The applied magnetic fields almost do not change the intensity of CDW and move the $T_{cdw}$, but shift the $T_{cant}$ to low temperatures, as described in the insets of Fig. \ref{1}(d)-\ref{1}(e). In other words, CDW in FeGe is robust against magnetic fields but the canting magnetic order can be tuned by fields. Such large divergency in $\rho_{ab}$ and $\rho_c$ suggests the different scattering mechanism for in-plane and out-of-plane transports. The anisotropic behaviors of transports are also observed in the magnetization data in Fig. \ref{1}(f)-\ref{1}(g). As $H \parallel$ \emph{ab}-plane, the $M_{ab}$ displays a linear dependence on the magnetic field up to 14 T with temperature range of 5 K to 240 K, but the $M_{c}$ for $H \parallel$ \emph{c}-axis undergoes a field-induced spin-flop transition at approximately 6 T below 100 K and 9 T above 120 K. A spin-flop transition generally occurs when applying a magnetic field is parallel to the moments along \emph{c}-axis (easy magnetization axis), it tends to rotate the magnetization perpendicular to the applied field. Instead of a saturated magnetization, the $M_{c}$ continues to linearly increase over the critical field, implying a higher polarization field in FeGe.

\begin{figure}
\includegraphics[angle=0,width=8cm,clip]{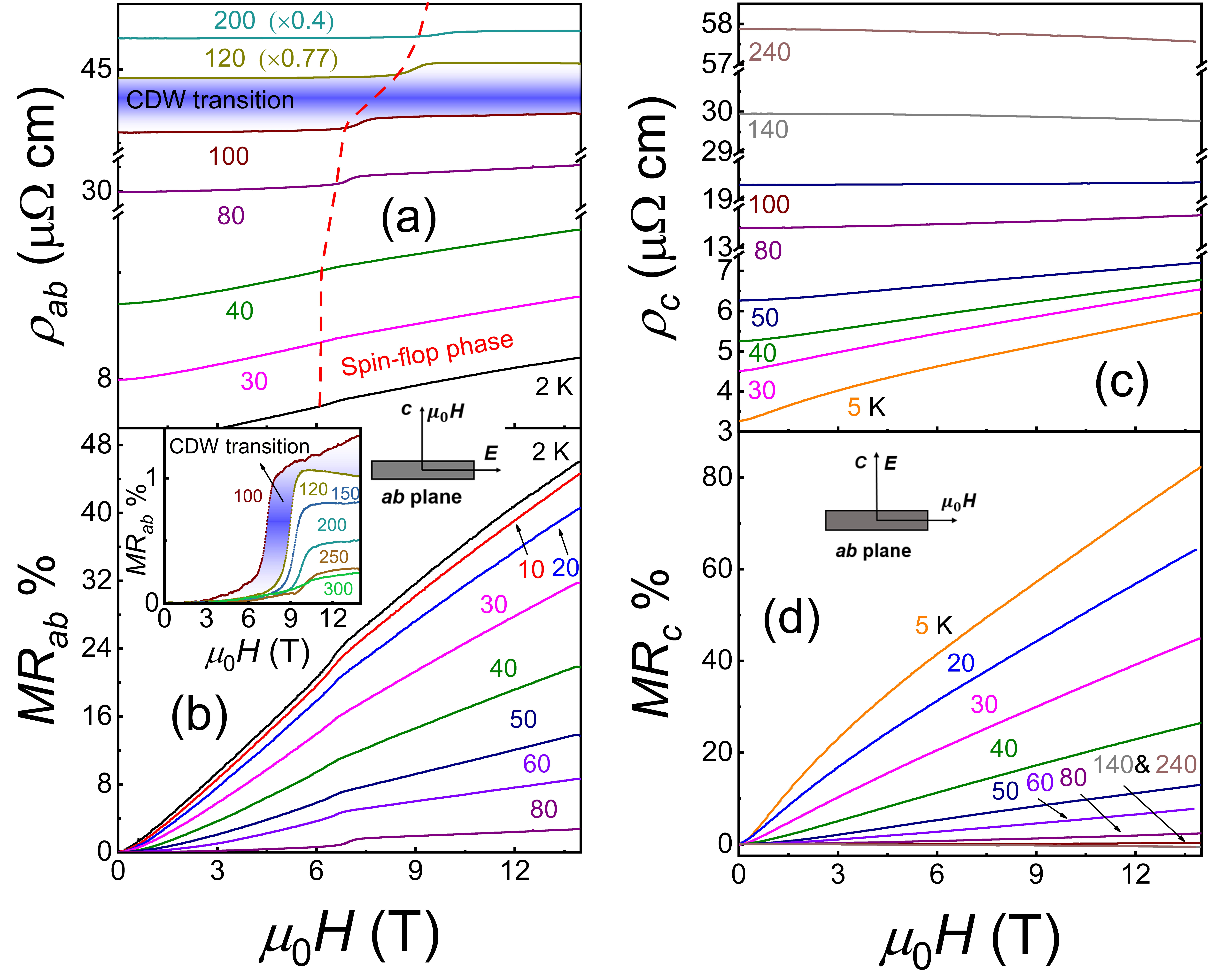}
\caption{(a)Resistivity $\rho_{ab}$ as a function of magnetic fields in the annealed FeGe and (b) its MR vs. fields \textit{H} $\parallel$ \textit{c}-axis $\perp$ \textit{I}. The inset shows the MR at high temperatures over 100 K. (c) Resistivity $\rho_{c}$ and (d) the MR vs. fields as \textit{H} $\parallel$ \textit{ab}-plane $\perp$ \textit{I}.}
\label{2}
\end{figure}
	
Applying a magnetic field into \textit{c}-direction and \textit{ab}-plane, the large different magnetic responses in $\rho_{ab}$ and $\rho_c$ are plotted in Fig. \ref{2}(a)-\ref{2}(d). Overall, both $\rho_{ab}$ and $\rho_c$ quickly increase with the increase of magnetic fields in the measured temperature range from 2 K to 240 K. In Fig. \ref{2}(a), $\rho_{ab}$ starts to increase linearly at low fields, followed by an obvious kink related to a field-induced spin-flop transition near 6 T (defined as $H_{sf}$). As \textit{H} $\geq$ $H_{sf}$, $\rho_{ab}$ exhibits the different characters with increasing temperature. It undergoes a crossover from a linear increase below $T_{cdw}$ to a constant above $T_{cdw}$ (see the inset of Fig. \ref{2}(b)). Meanwhile, the $H_{sf}$ remains almost unchange
but amplitude of this transition becomes more prominent with warming temperature to 100 K. But above $T_{cdw}$ the $H_{sf}$ is shifted to higher fields and its amplitude starts to weaken. Such feature implying the strongly correlation between charge and AFM order is not yet observed in other kagome magnets so far. Now turning to $\rho_{c}$ in Fig. \ref{2}(c), the spin-flop transition is of absence and it shows the linear field-dependent in the whole temperature regime. Its magnetoresistivity (MR)in Fig. \ref{2}(d) reaches 80$\%$ at 5 K and 14 T, which is about twice times larger than the MR value of 43$\%$ at 2 K for $\rho_{ab}$ in Fig. \ref{2}(b). Employing the Kohler rule to examine the MR can reflect the information of the band structure like Fermi surface nesting, as demonstrated in other CDW materials\cite{PRLTiSe2,TiSe2}. The nonlinear $(B/\rho_0)^2$ curves in $\rho_{c}$ do not scale with each other (See detailed information in SI) and the Kohler slope shows a monotonic variation with temperature across the $T_{cdw}$, in contrast sharply to those observed in TiSe$_2$ with a minimum at $T_{cdw}$, implying a distinct origin of the MR in the FeGe system.
	
\begin{figure*}
\includegraphics[angle=0,width=16cm,clip]{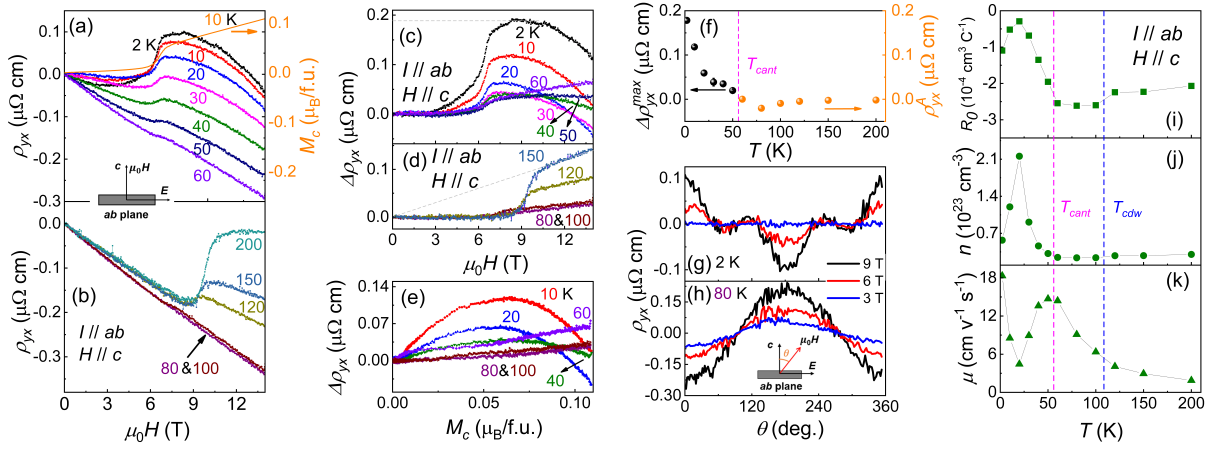}
\caption{ For \textit{H} $\parallel$ \textit{c}-axis, \textit{I} $\parallel$ \textit{ab}-plane, (a)-(b) Hall resistivity $\rho_{yx}$ vs. magnetic fields in the annealed FeGe with temperature range from 2 K to 200 K. As a compare, the magnetization at 10 K is also plotted in figure 4(a). (c)-(d) The obtained $\Delta\rho_{yx}$ as a function of magnetic fields below 200 K. (e) Magnetization dependence of $\Delta\rho_{yx}$ below $T_{cdw}$ [The magnetization data used for making this plot are shown in figure 2f]. (f) Temperature dependent of the maximum $\Delta\rho^{max}_{yx}$ of in-plane current and the $\rho^A_{yx}$ of out-of-plane current. (g)-(h) Hall resistivity $\rho_{yx}$ at 2 K and 80 K as a function of angle at different fields. The inset defines the rotation angle $\theta$. (i)-(k) Hall coefficient (i), carrier densities (j), and mobilities (k) as functions of temperature with current along \textit{ab}-plane, respectively. }
\label{3}
\end{figure*}
	
We then examine the Hall resistivity as a function of magnetic fields (\textit{H} $\parallel$ \textit{c}-axis and \textit{I} $\parallel$ \textit{ab}-plane) in Fig. \ref{3}(a)-\ref{3}(b). Consistent with the $\rho_{ab}$, Hall resistivity $\rho_{yx}(H)$ also exhibits the field-induced spin-flop transition at $H_{sf} \sim 6 $  T below 100 K. It does not shift significantly but weakens considerably with temperature warming up to 100 K. At higher temperatures, the transition suddenly shifts to 9 T, and its amplitude intensifies significantly at 200 K. However, at a low temperature of 2 K, $\rho_{yx}(H)$ follows nearly a linear field-dependent at low fields, denoting Hall coefficient $R_0 $ and carrier density $n$ of $\sim 1 \times10^{-10}$~m$^3$/C  and $\sim 6\times 10^{22}$ cm$^{-3}$, respectively, as illustrated in Fig. \ref{3}(i-j). Beyond $H_{sf}$, a broad hump in $\rho_{yx}(H)$ emerges and persists until $T_{cdw}$, finally recovers an approximately linear field-dependent at $T_{cdw}$. Until 200 K, $\rho_{yx}(H)$ tends to hold a constant over $H_{sf}$. Such ``anomalous" Hall effect observed in our annealed FeGe crystal is very distinct from the findings in Ref.\cite{FeGe}, indicating a different origin.

Note Hall resistivity $\rho_{yx}$ does not exactly follow the magnetization at a low temperature of 10 K (more magnetization data in Fig. \ref{1}(g)), as illustrated in Fig. \ref{3}(a), possibility involving a topological Hall effect (THE) observed in other magnetic materials with noncollinear/noncoplanar spin texture (NSF) such as MnBi$_2$Te$_4$ \cite{MnBiTe}, CeAlSi/Ge \cite{CeAlGe,CeLuo}, Fe$_3$GeTe$_2$ \cite{FeGeTeTHE}. Considering the CDW transition and the canted AFM order, we can divide the Hall effect into three regimes$:$ (a) in the low temperature region below $T_{cant}$, a large topological Hall effect dominates due to the noncollinear/coplanar spin texture in the canted AFM state, similar to that in MnBi$_2$Te$_4$ \cite{MnBiTe}; (b) in the intermediate region from $T_{cant}$ to $T_{cdw}$, a weak anomalous Hall effect is detected, which is attributed to the competition of canting AFM fluctuation and CDW order; (c) in the high temperature range over $T_{cdw}$, the field-induced AHE associated with spin-flop transition is hard to be extracted because of the collinear A-type AFM state. We thus can obtain THE and AHE by subtracting the normal Hall effect at low fields using the formula: $ \rho_{yx} = R_0 \mu_{0}H +\rho^A_{yx}+ \rho^T_{yx}$. The obtained $\Delta \rho_{yx} = \rho_{yx} - R_0 \mu_{0}H$ (Fig. \ref{3}(c) and \ref{3}(d)) shows a broad hump at high fields, reaching a maximum of 0.2 $\mu \Omega$~cm. Increasing temperature to $T_{cant}$, $\Delta \rho_{yx}$ gradually decreases with the weakening of the hump, instead it holds a constant at high fields for 50 K. In contrast, above $T_{cdw}$ the $\Delta \rho_{yx}$ exhibits a significant jump at $H_{sf}$, followed by a linear field-dependent in Fig. \ref{3}(d), consistent with the magnetization data (see Fig. \ref{1}(g)). This THE can be demonstrated by plotting $\Delta \rho_{yx}$ as a function of magnetization $M_c$ in Fig. \ref{3}(e). The $\Delta \rho_{yx}(M)$ exhibits significant humps associated with THE in the CAFM state, which is attributed to the noncollinear spin texture caused by the field-induced spin-flop transition. Similar observations of spin-flop transition and THE are also evident in the AFM MnBi$_2$Te$_4$ sample\cite{MnBiTe}. We thus plot the in-plane Hall resistivity $\Delta\rho^{max}_{yx}$ as a function of temperature in Fig. \ref{3}(f). The maximum of topological hall resistivity $\Delta\rho^{max}_{yx}$ decreases monotonically with increasing temperature to $T_{cant}$. Above $T_{cant}$, the THE is of absence, meanwhile the anomalous Hall effect $\rho^A_{yx}$ is hardly distinguished though the $\rho_{yx}$ shows an obvious jump at $H_{sf}$ because of spin-flop transition. The results indicate that the THE can be attributed to the noncollinear/non-coplanar spin texture when the sample enters into the canting AFM states along $\textit{c}$-axis, similar to reports in other topological magnets MnBi$_2$Te$_4$\cite{MnBiTe}, CeAlSi/Ge\cite{CeAlGe,CeLuo}.

To further clarify the THE in the non-collinear AFM state of FeGe, we present the Hall resistivity as a function of the angle between $H$ and the \emph{c}-axis in Fig. \ref{3}(g)-\ref{3}(h). At 2 K and 3 T, the $\rho_{yx}$ remains a detectable small value. After the spin-flop transition, when the applied magnetic fields exceed 6 T, it can be found that the value of $\rho_{yx}$ decreases quickly with the $H$ away from the \emph{c}-axis. In a specific angle range from 60$^\circ$ to 120$^\circ$, the  $\rho_{yx}$ is almost completely vanished. This means that the THE in the approximate in-plane setup does not exist. In contrast, temperature is up to 80 K above the $T_{cant}$, the $\rho_{yx}$ at 9 T shows a smooth evolution and follows a cos$\theta$ relationship with angle. The similar observation has been previously documented in a frustrated Gd$_2$PdSi$_3$ system with skyrmionic lattice\cite{GdPdSi} and an AFM topological insulator MnBi$_4$Te$_7$ system\cite{MnBITe147}. Temperature dependence of Hall coefficient, carriers density, and mobilities is shown in Fig. \ref{3}(i)-\ref{3}(j). The in-plane $R^{ab}_{0}$ is negative, shows a slight drop at $T_{cdw}$ but sharply decreases below $T_{cant}$. The in-plane carrier density, $n_{ab}$, increases substantially below $T_{cant}$ with a comparable value of $\sim 10^{22}$~cm$^{-3}$, similar to the reported results\cite{FeGe}. The mobility $\mu_{ab}$ slightly increases with decreasing temperature, followed by a broad peak around $T_{cant}$.

\begin{figure}
\includegraphics[angle=0,width=8cm,clip]{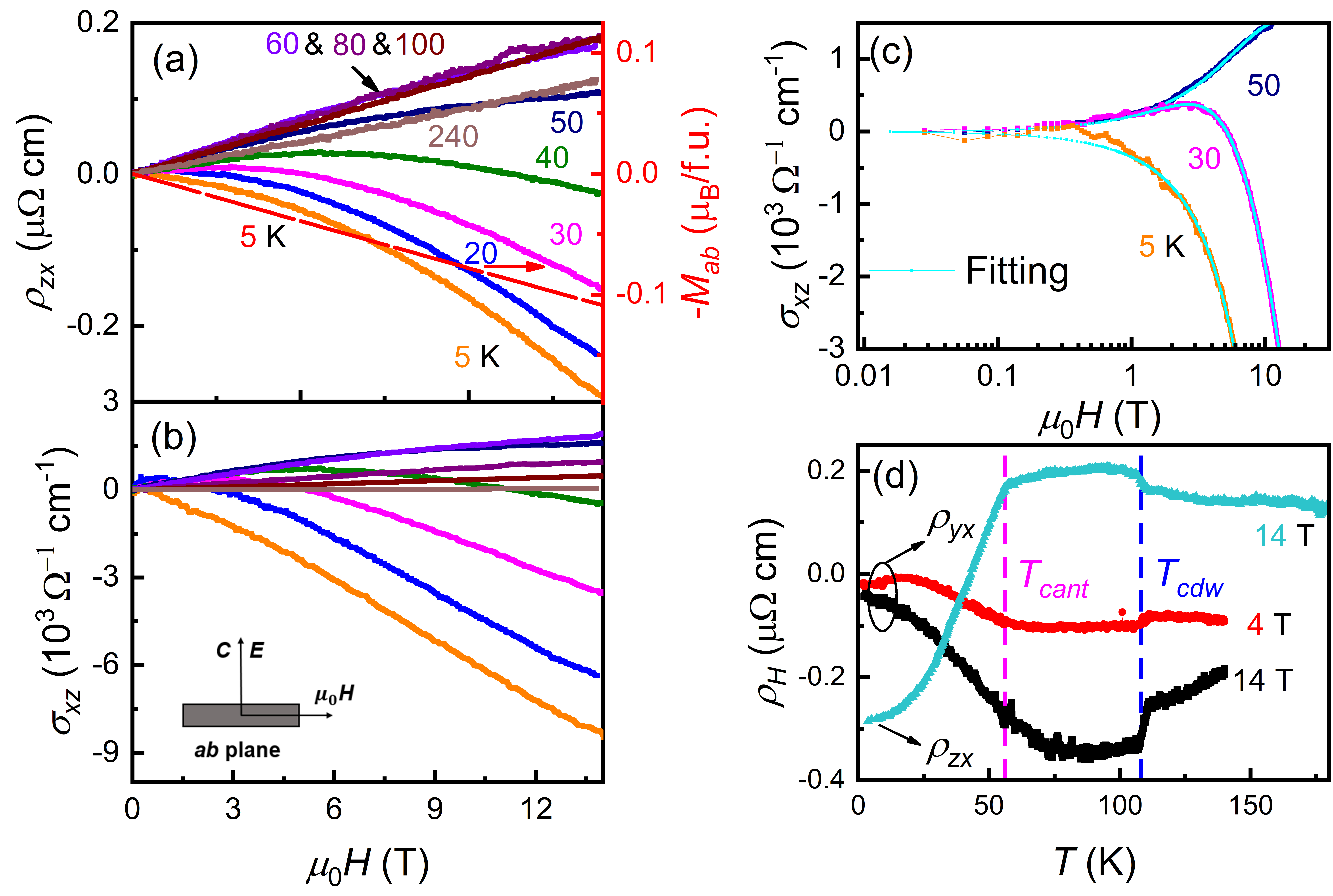}
\caption{ The Hall resistivity $\rho_{zx}$ (a) and its conductivity (b) as a function of magnetic fields as the current flows along \emph{c}-direction and the magnetic field parallels to \emph{ab}-plane. (c) The fitting results of two-band modal at several selected temperatures. (d) Temperature dependence of Hall resistivity $\rho_{yx}$ and $\rho_{zx}$ at 4 T and 14 T. }
\label{4}
\end{figure}

As \textit{H} $\parallel$ \textit{ab}-plane and \textit{I} $\parallel$ \textit{c}-direction, the out-of-plane Hall effect shows large differences in Fig. \ref{4}(a). It is found that Hall resistivity $\rho_{zx} (H)$ obviously bends at low fields but obeys a linear behavior at high fields for 5 K, in conflict with the magnetization with the linear field-dependent along \textit{ab}-plane (Fig. \ref{1}(f)). This similar feature has been observed in the nonmagnetic kagome compound KV$_3$Sb$_5$\cite{KVSb} and ScV$_6$Sn$_6$\cite{ScVSn}, which is ascribed to either an AHE or a two-band normal Hall effect (NHE). The Hall resistivity curves, however, persist and remain visible until 100 K at which $\rho_{zx}(H)$ recovers a linear field-dependent, accompanying with the slope of $\rho_{zx}(H)$ changing from negative to positive with temperature. We attempted to fit the hall data using two-band modal\cite{PRLTiSe2}. The Hall conductivity $\sigma_{xz}$ in Fig. \ref{4}(b) is thus calculated by inverting the resistivity matrix, $\sigma_{zx} = -\rho_{zx}/({\rho_{zx}}^2+\rho_{zz}^2)$. Above 20 K, the two-band modal fittings in $\sigma_{xz}$ (Fig. \ref{4}(c)) match very well, but for 5 K it seriously violates the experimental data at low fields. Considering the linear MR behavior in Fig. \ref{2}(d), we found that the fitting simultaneously of both MR and Hall data using the two-band model is hardly possible, suggesting other mechanisms such as skew/side-jump scattering or charge order induced unconventional Hall effect\cite{ScVSn,KVSb} may be involved in the present system. Fig. \ref{4}(d) plots temperature dependence of the anisotropic Hall effect. As $B$ = 14 T, $\rho_{yx}(T)$ is negative while $\rho_{zx}(T)$ is positive above $T_{cant}$, meanwhile $\rho_{yx}(T)$ shows a more prominent jump at $T_{cdw}$ than that of $\rho_{zx}(T)$, implying that the charge order can effectively enhance the Hall resistivity. Entrance of the canted AFM state below $T_{cant}$, the $\rho_{zx}(T)$ starts to sharply drop to zero and then continues to increase with the sign-reversal, but the decreasing $\rho_{yx}(T)$ remains negative until the measured lowest temperature. The divergency behaviors suggest the different origin of anomalous Hall effect, and also imply that the effect of magnetic ordering on $\rho_{yx}(T)$ plays an important role. These facts also reveal the strong relationship among magnetism, charge order and topology in the kagome magnets FeGe.

\begin{figure}
\includegraphics[angle=0,width=8cm,clip]{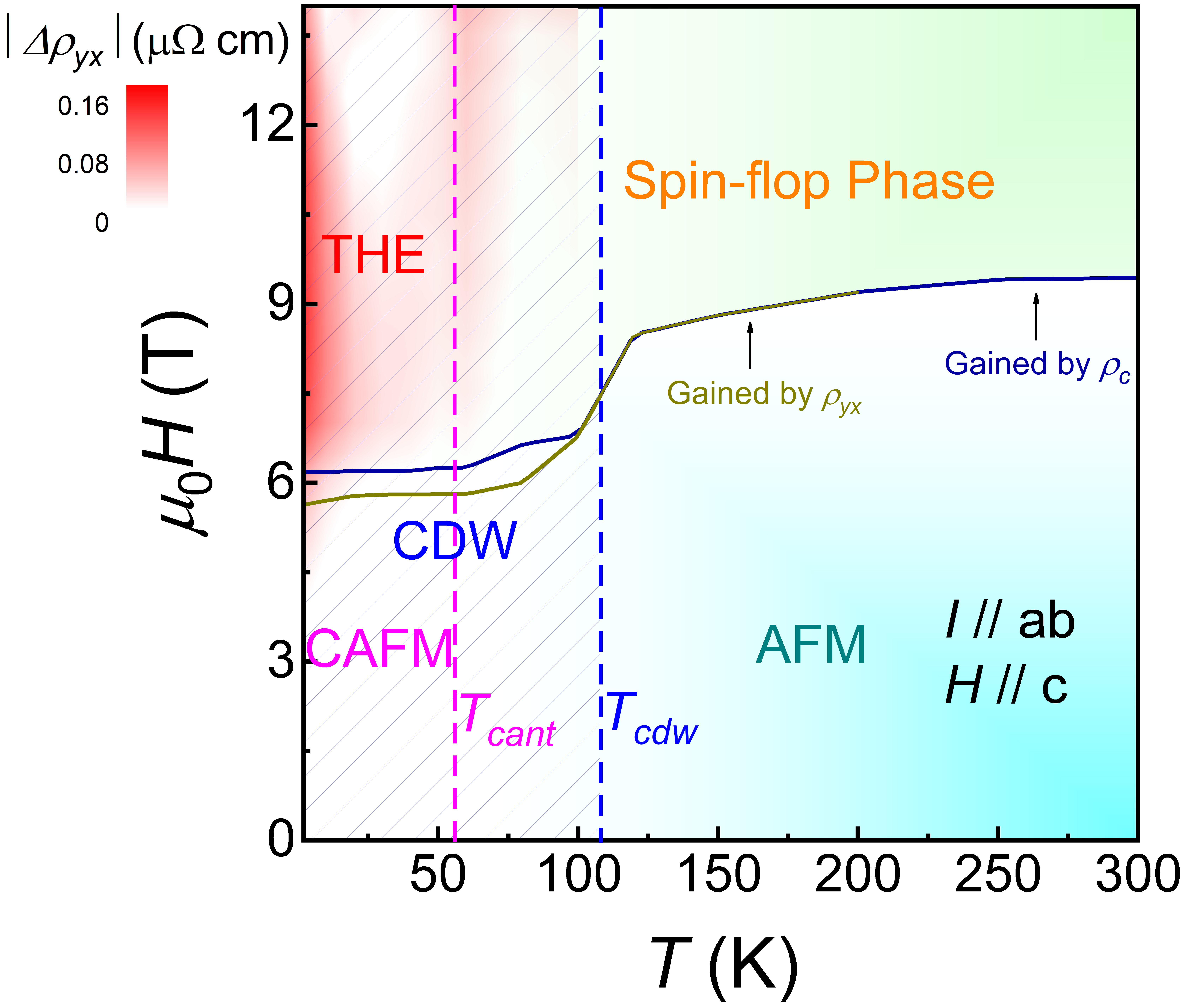}
\caption{ Magnetic Phase diagram of FeGe and contour plot of topological Hall resistivity as a function of the external magnetic field and temperature as \textit{I} $\parallel$ \textit{ab}-plane, \textit{H} $\parallel$ \textit{c}-axis.}
\label{5}
\end{figure}
	
According to the reported data and our measured results, magnetic phase diagram and topological Hall resistivity as a function of external magnetic field and temperature are established in the Fig. \ref{5} as \textit{H} $\parallel$ \textit{c}-axis and \textit{H} $\parallel$ \textit{ab}-plane (see Figure S4 in SI). The phase diagram does not exactly follow the previous result\cite{FeGe}. The applied magnetic fields induce the spin-flop transition at $H_{sf}$, but the $H_{sf}$ undergoes a sharp crossover at $T_{cdw}$, where it starts to remain $\sim$ 6 T at low temperatures but suddenly jumps to a constant $\sim$ 9 T around $T_{cdw}$. The lower $H_{sf}$ happens to appear below $T_{cdw}$, indicating a strong correlation between charge order and magnetism in FeGe. Below $T_{cant}$, the magnetic fields do not change the CDW order, but induce a spin-flop transition. This spin-flop process may flop from easy-axis to the hard axis and transform the canted double-cone AFM into the non-collinear AFM phase, resulting in the appearance of a large THE (the contour plot). Above $T_{cant}$, the spins tend to become parallel, leading to the suppression of THE.
	
Indeed, theoretical model \cite{Ong} has suggested the intrinsic AHE is related to Berry phase concept. In some magnetic or Dirac/Weyl materials such as the Fe film \cite{PhysRevLett.99.086602}, the kagome-ferromagnet Co$_3$Sn$_2$S$_2$ \cite{MPI,wangRM,CoSnSLi}, and the nodal-line ferromagnet Co$_2$MnGa\cite{CoMnGa}, the non-trivial Berry curvature caused by the magnetization results in the observation of AHE. Another AHE is caused by the topological nontrivial spin structures derived by geometrical frustration or Dzyaloshinskii–Moriya interaction, termed as topological Hall effect. In the topological magnets with noncollinear/noncoplanar spin textures, the itinerant electrons with a finite scalar spin chirality $\chi_{ijk}=S_i\cdot(S_j\times S_k)$ can acquire the real-space Berry phase, leading to the large THE like MnGe \cite{PRLMnGe}, Mn$_3$Sn \cite{PRBMnSn3} , Mn$_2$PtIn \cite{PhysRevB.102.014449} and CoNb$_3$S$_6$\cite{takagi2023spontaneous}. In addition to those studies, charge order is also believed to dominate the AHE in the nonmagnetic kagome AV$_3$Sb$_5$ and ScV$_6$Sn$_6$ compounds\cite{ScVSn,KVSb,KVSbChiral,PRBCsVSb} because of the loop-current-induced the broken time reversal symmetry. Here, the canted double-cone AFM structure below $T_{cant}$ in the FeGe system rotates an inter-layer turn angle, giving birth to a basal-plane moment component$: M_{ij} = M_0 \mathrm{sin}(\overrightarrow{q}\cdot \overrightarrow{r_i}+ \phi)$, where $M_0$ is the maximum moment component, $\overrightarrow{q}$ is the modulation vector and $\phi$ an arbitrary phase. This moment component $M_{ij}$ increases with temperature decreasing due to the increasing cone half-angle in the double-cone AFM structure. Therefore, the observed THE can be attributed to the Berry phase resulting from the non-collinear spin textures in the canted AFM phase, similar to the reported in MnBi$_2$Te$_4$\cite{MnBiTe}, Gd$_2$PdSi$_3$\cite{GdPdSi} and MnBi$_4$Te$_7$ \cite{MnBITe147}.

\section{IV. CONCLUSION}
	
In summarize, we investigate the unexpected anisotropic transport properties observed in the annealed FeGe crystals, which feature a crystal structure composed of stacked Fe kagome planes along the \textit{c}-axis. All measurements including magnetic susceptibility, (magneto) resistivity and Hall effect under both in-plane and out-of-plane magnetic fields reveal significant anisotropic characteristics, implying the different transport mechanisms for the two directions. Notably, the abnormal anisotropic resistivity is unusual, with the in-plane resistivity $\rho_{ab}$ being approximately three times larger than $\rho_c$ at room temperature, likely due to the anisotropic band structures. Importantly, a large THE can be observed during this spin-flop process in the canted AFM phase due to the non-collinear/coplanar spin texture as \textit{H} $\parallel$ \textit{c}-axis, while as \textit{H} $\parallel$ \textit{ab}-plane, the Hall resistivity exhibits a deviation from the linear field-dependent despite the absence of net magnetization. The phenomenon remains unclear, which may be explained by various models including the two-band modal, the charge order with a chiral loop-current effect or the topological Dirac points. However, the anisotropic transport behaviors in FeGe cannot solely be attributed to its crystal structures or canted-AFM configurations. Other factors, such as anisotropic band structures, nontrivial spin textures and the interplay of CDW, magnetism and topology, must also be considered. Our findings not only deepen the comprehension of the origin of the anisotropic AHE but also highlight the intricate relationships among magnetism, charge order, and topology in the kagome metal FeGe.

	\section*{ACKNOWLEDGEMENTS}
This research was supported in part by the Nature Science funding (NSF) of China (under Grants No. U1932155, 12374116, 12204298, 12274019), Y. Li acknowledges the Hangzhou Joint Fund of the Zhejiang Provincial Natural Science Foundation of China (under Grants No. LHZSZ24A040001). H. Guo was funded by the Guangdong Basic and Applied Basic Research Foundation (Grant No. 2022B1515120020).
	
	\noindent\
	\\
	* \verb|Equal contributions.|\\
	\dag  \verb|jinke_bao@hznu.edu.cn|\\
	\ddag \verb|yklee@hznu.edu.cn|\\

	\bibliographystyle{scpma}
	\bibliography{FeGe}
	
\end{document}